\documentclass[conference]{IEEEtran}
\IEEEoverridecommandlockouts

\usepackage{textcomp}

\usepackage{cite} 

\usepackage[caption=false,font=footnotesize]{subfig}	
\usepackage{graphicx}
\usepackage{float}

\usepackage{algorithm}
\usepackage{algorithmicx}
\usepackage{algpseudocode}

\usepackage{amsmath,amsfonts,amssymb,amsthm}
\usepackage{mathtools}
\usepackage{mathrsfs}	
\usepackage{subdepth}	

\usepackage{array}	
\usepackage{arydshln}   
\usepackage{mathtools}
\usepackage{tabularx}   

\usepackage{hyperref}
\usepackage[sort,compress,capitalise]{cleveref}	
\usepackage{url}

\usepackage[dvipsnames]{xcolor} 
\usepackage{xspace} 

\usepackage{tikz}
\usetikzlibrary{fit}
\usetikzlibrary{matrix,decorations.pathreplacing, calc, positioning}
\usetikzlibrary{babel}  
\usepackage[straightvoltages,nooldvoltagedirection]{circuitikz}
\usetikzlibrary{calc,shapes.geometric}
\usepackage{scalerel}

{
\theoremstyle{plain}

}

{
\theoremstyle{definition}

}

\crefname{Hypothesis}{Hyp.}{Hyps.}
\Crefname{Hypothesis}{Hyp.}{Hyps.}

\crefname{Lemma}{Lemma}{Lemmata}
\Crefname{Lemma}{Lemma}{Lemmata}

\crefname{Definition}{Def.}{Defs.}
\Crefname{Definition}{Def.}{Defs.}

\newcommand{\CIGRE}{CIGR{\'E}\xspace}

\newcommand{\DFT}[1][]{DFT\xspace}  

\newcommand{\RMS}{RMS\xspace}   

\newcommand{\TE}[1][]{\textup{TE#1}\xspace}   

\newcommand{\AC}{AC\xspace} 
\newcommand{\DC}{DC\xspace}	

\newcommand{\DER}[1][]{DER#1\xspace}    
\newcommand{\CIDER}[1][]{CI\DER[#1]}    
\newcommand{\NIC}[1][]{NIC#1\xspace}         

\newcommand{\TDS}{TDS\xspace}	
\newcommand{\HPF}{HPF\xspace}	

\newcommand{\KPI}[1][]{KPI#1\xspace}    

\newcommand{\Abs}[1]{\left|#1\right|}

\newcommand{\Set}[1]{\mathcal{#1}}

\newcommand{\derivative}[2]{\frac{\partial #1}{\partial #2}}

\newcommand{\spt}{\sigma}

\newcommand{\Y}{\mathbf{Y}}     

\newcommand{\phases}{\Set{P}}
\newcommand{\nodes}{\Set{N}}

\newcommand{\phsABC}{\texttt{ABC}}

\newcommand{\cmpDQ}{\texttt{DQ}}

\newcommand{\formers}{\Set{S}}
\newcommand{\followers}{\Set{R}}

\newcommand{\VT}{\mathbf{v}}
\newcommand{\IT}{\mathbf{i}}

\def\BibTeX{{\rm B\kern-.05em{\sc i\kern-.025em b}\kern-.08em
    T\kern-.1667em\lower.7ex\hbox{E}\kern-.125emX}}
\begin{document}

\title{Harmonic Power-Flow Study of Hybrid \AC/\DC Grids with
    Converter-Interfaced Distributed Energy Resources
\thanks{This work was funded by the Schweizerischer Nationalfonds (SNF, Swiss National Science Foundation) via the National Research Programme NRP~70 ``Energy Turnaround'' (projects nr. 173661 and 197060) and by the Deutsche Forschungsgemeinschaft (DFG, German Research Foundation) via the Priority Programme DFG~SPP~1984 ``Hybrid and Multimodal Energy Systems'' (project nr. 359982322).}
}

\author{\IEEEauthorblockN{Johanna~Kristin~Maria~Becker, Yihui~Zuo, Mario~Paolone}
\IEEEauthorblockA{\textit{Distributed~Electrical~Systems~Laboratory} \\
\textit{{\'E}cole Polytechnique F{\'e}d{\'e}rale de Lausanne}\\
Lausanne, Switzerland \\
\{johanna.becker, yihui.zuo, mario.paolone\}@epfl.ch}
\and
\IEEEauthorblockN{Andreas~Martin~Kettner}
\IEEEauthorblockA{
\textit{PSI~NEPLAN~AG}\\
Küsnacht (Zürich), Switzerland \\
andreas.kettner@neplan.ch}
}

\maketitle

\begin{abstract}
    As the share of \emph{Converter-Interfaced Distributed Energy Resources} (\CIDER[s]) in power distribution systems increases, hybrid \AC/\DC distribution systems are drawing more interest.
    Indeed, \CIDER[s] usually rely on \DC power and hybrid \AC/\DC grids with few \emph{Network-Interfacing Converters} (\NIC[s]) are a promising solution to decrease installation costs and conversion losses compared to a pure \AC grid.
    However, interactions between the \AC and \DC subsystem of a hybrid \AC/\DC grid can lead to undesirable amplification and propagation of harmonics.
    The authors of this paper have recently proposed a \emph{Harmonic Power-Flow} (\HPF) method that accurately represents the \AC power flows including the coupling between different harmonics.
    The \HPF framework is formulated through the mismatch equations of the nodal equations between the grid and the resource models and solved by a Newton Raphson algorithm.
    This paper updates the \HPF method to model hybrid \AC/\DC grids interconnected through \NIC[s].
    To this end, the model of the \NIC[s] and the resulting coupling between the \AC and \DC subsystems is included in the mismatch equations and the Jacobian matrix of the algorithm.
    The updated \HPF method is applied to a typical hybrid \AC/\DC grid, and its accuracy is validated through detailed time-domain simulations with Simulink.
\end{abstract}


\section{Introduction}



Modern power distribution systems are experiencing a large-scale integration of distributed energy resources.
Since the majority of these resources rely on \DC power, they need to be interfaced to the \AC grid through \AC/\DC converters.
Due to the potential to exploit synergies of the \DC components of such \emph{Converter-Interfaced Distributed Energy Resources} (\CIDER[s]), the concept of \DC~microgrids has emerged.
\DC~microgrids are promising in terms of eliminating redundant converter stages along with the associated installation cost and energy losses~\cite{Jrn:2014:Eghtedarpour}.
Since \AC grids are still dominant in today’s power system, it is not expected that they will be replaced by \DC grids entirely.
Instead, so-called hybrid \AC/\DC grids are envisioned as a possible architecture to profit from the advantages of both types of systems~\cite{Jrn:2014:Nejabatkhah}.
In a hybrid \AC/\DC grid, the \AC and \DC subsystems are interconnected through so-called \emph{Network-Interfacing Converters} (\NIC[s]).
The interconnection of \AC and \DC grids can help to augment the reliability and stability of the entire system if adequate control schemes are used~\cite{Cnf:2011:Ambia}.

The presence of numerous \CIDER[s] can cause high levels of harmonic distortions in power distribution systems~\cite{Jrn:PSE:PEC:2004:Enslin}.
Moreover, when interfacing subsystems composed of entire \AC and \DC grids, significant harmonic propagation between them is expected to occur~\cite{Jrn:2019:Nejabatkhah}.
In order to mitigate undesired amplification of harmonics in such cases, the appropriate design and tuning of the controllers of \CIDER[s] and \NIC[s] is crucial~\cite{Jrn:2019:Nejabatkhah}.
To this end, accurate methods for the computation and analysis of harmonics and their propagation through \CIDER[s], \NIC[s], and the grid are required.



Power-flow methods for the analysis of hybrid \AC/\DC grids have been studied extensively for transmission systems.
In the literature, these methods are classfied as \emph{sequential} or \emph{unified} algorithms.
A sequential algorithm solves the \AC and \DC equations separately~\cite{Jrn:2012:Beerten}, while a unified algorithm solves the complete set of equations jointly~\cite{Jrn:2013:Baradar}. 
In \cite{Cnf:2010:Beerten} and \cite{Jrn:2013:Baradar}, power-flow analyses are performed for \AC grids including HV\DC links by means of a unified and sequential approach, respectively.
A comparison of a unified and a sequential \emph{Harmonic Power-Flow} (\HPF) method for hybrid \AC/\DC grids including a single HV\DC link is performed in \cite{Jrn:1999:Smith}.

A power-flow method for hybrid \AC/\DC grids on distribution system level is proposed in~\cite{Jrn:2018:Murari}.
The method is formulated in a unified fashion and is capable of analysing entire \DC grids as opposed to single HV\DC links.
To the best of the authors' knowledge, no generic and modular \HPF study exists that allows the analysis of entire hybrid \AC/\DC grids while modelling in detail the harmonics in \CIDER[s] and \NIC[s].
%


The authors of this paper have recently proposed a framework for the \HPF study of polyphase grids with a high share of \CIDER[s]~\cite{jrn:2020:kettner-becker:HPF-1,jrn:2020:kettner-becker:HPF-2,jrn:2022:becker}.
This method is generic, modular and accurate and includes the coupling between harmonics.
The models are described by linear time-periodic state-space models, which are transformed to the frequency domain by means of the Fourier analysis and Toeplitz matrices.
The \HPF problem is formulated based on the closed-loop transfer functions of the \CIDER[s] and the hybrid nodal equations of the grid, and solved using the Netwon Raphson method.
The \CIDER model proposed in~\cite{jrn:2022:becker} represents both the \AC- and \DC-side components and their controllers in detail.


This paper extends the \HPF algorithm to be applicable to hybrid \AC/\DC grids.
To this end, the model of the \NIC[s] which interfaces the \AC and \DC grids is derived as an extension of the generic \CIDER model in \cite{jrn:2022:becker}.
Furthermore, the structure of the \HPF method and the Jacobian matrix required for the Newton-Raphson algorithm are updated from the previous versions.
More precisely, the Jacobian matrix is built through combination of the Jacobian matrices of the individual subsystems along with the respective coupling terms which are due to the \NIC[s].

The rest of this paper is structured as follows.
\Cref{sec:NICs} gives a short summary of the time-domain model of a \NIC and how it is obtained from the one of a \CIDER as proposed in \cite{jrn:2022:becker}.
In \cref{sec:Resources} the node partition known from previous papers is refined for hybrid \AC/\DC grids.
Furhtermore, the harmonic-domain grid responses of the resources are shortly recalled from previous papers and introduced for the case of the \NIC[s].
\Cref{sec:Algorithm} shows the changes in the \HPF algorithm for hybrid \AC/\DC grids compared to the previous versions.
The \HPF method for hybrid \AC/\DC grids is validated in \cref{sec:val-sys} and the conclusions are drawn in \cref{sec:Conclusions}.

\section{Model of a Network-Interfacing Converter}
\label{sec:NICs}

\NIC[s] interconnect the \AC and \DC subsystems of a power grid.
The structure of a \NIC is similar to the one of a \CIDER proposed in Section~IV~of~\cite{jrn:2022:becker}.
Like a \CIDER, a \NIC consists of power hardware and control software.

\begin{figure}[htbp]
	\centering
    {

\ctikzset{bipoles/length=1cm}
\ctikzset{bipoles/diode/height=.2}
\ctikzset{bipoles/diode/width=.15}
\tikzstyle{block}=[rectangle, draw=black,fill=white, minimum size=8mm, inner sep=0pt]
\tikzstyle{dot}=[circle, draw=black, fill=black, minimum size=2pt, inner sep=0pt]
\tikzstyle{measurement}=[rectangle,draw=black,minimum size=1mm,inner sep=0pt]
\tikzstyle{signal}=[-latex]

\def\transform#1#2
{%
\begin{scope}[shift={#2}]
    \node[block] (#1) at (0,0) {};
    \draw (#1.south west) to (#1.north east);
    \node at ($(#1.north west)+(0.3,-0.15)$) {\scriptsize$\phsABC$};
    \node at ($(#1.south east)+(-0.2,0.15)$) {\scriptsize$\cmpDQ$};
\end{scope}
}

\definecolor{blue}{rgb}{0.74, 0.83, 0.9}
\definecolor{green}{rgb}{0.78, 0.9, 0.82}
\definecolor{red}{rgb}{0.97, 0.71, 0.67}
\definecolor{yellow}{rgb}{1.0, 0.99, 0.82}

\footnotesize

\begin{circuitikz}
    
    \def\x{1.2}
    \def\y{1.2}
    
    
    
    \node[rectangle,draw=black,fill=white,minimum height=15mm,minimum width=12mm,inner sep=0pt] (C) at (0,0) {};
    \node[nigbt,scale=0.8] (IGBT) at (C) {};
    \draw (IGBT.E)++(0,0.1) -- ++(0.3,0) to[D*] ($(IGBT.C)+(0.3,-0.1)$)   -- ++(-0.3,0);

    \coordinate (AN) at ($(C.east)-(0*\x,0.5*\y)$);
    \coordinate (AP) at ($(C.east)+(0*\x,0.5*\y)$);
    
    \coordinate (FALN) at ($(AN)+(0.25*\x,0)$);
    \coordinate (FALP) at ($(FALN)+(0,\y)$);
    \coordinate (FARN) at ($(FALN)+(0.65*\x,0)$);
    \coordinate (FARP) at ($(FARN)+(0,\y)$);
    
    \coordinate (FGLN) at ($(FARN)+(0.65*\x,0)$);
    \coordinate (FGLP) at ($(FGLN)+(0,\y)$);
    \coordinate (FGRN) at ($(FGLN)+(0.65*\x,0)$);
    \coordinate (FGRP) at ($(FGRN)+(0,\y)$);
    
    \coordinate (GN) at ($(FGRN)+0.5*(\x,0)$);
    \coordinate (GP) at ($(GN)+(0,\y)$);

    \draw (FALN) to[open] (FALP);
    
    \draw (AN) to[short] (FALN);
    \draw (AP) to[short] (FALP);

    \draw (FALN) to[short] (FARN);
    \draw (FALP) to[inductor] (FARP);
    
    \draw (FARN) to[short] (FGLN);
    \draw (FARP)
        to[short] ($0.5*(FARP)+0.5*(FGLP)$) 
        to[short] (FGLP); 
    
    \draw ($0.5*(FARN)+0.5*(FGLN)$) to[capacitor,*-*,fill=white] ($0.5*(FARP)+0.5*(FGLP)$); 
    
    \draw (FGLN) to[short] (FGRN);
    \draw (FGLP) to[inductor,fill=white] (FGRP); 
    
    \draw (FGRN) to[short,-o] (GN);
    \draw (FGRP) to[short,-o,i=$\IT^{\AC}(t)$] (GP);
    
    \draw (GN) to[open,v_=$\VT^{\AC}(t)$] (GP);
    
    
    
    \coordinate (DN) at ($(C.west)-(0*\x,0.5*\y)$);
    \coordinate (DP) at ($(C.west)+(0*\x,0.5*\y)$);
    
    \coordinate (FDRN) at ($(DN)-(0.5*\x,0)$);
    \coordinate (FDRP) at ($(FDRN)+(0,\y)$);
    \coordinate (FDLN) at ($(FDRN)-(0.5*\x,0)$);
    \coordinate (FDLP) at ($(FDLN)+(0,\y)$);
    
    \draw (DN) to[short] (FDRN);
    \draw (FDRP) to[short] (DP); 

    \draw ($(FDRN)$) to[capacitor,*-*,fill=white] ($(FDRP)$); 

    \draw (FDRN)
        to[short,-o] (FDLN);
    \draw (FDRP)
        to[short,-o,i_=$i^{\DC}(t)$] (FDLP);
        
    \draw (FDLN) to[open,v^=$v^{\DC}(t)$] (FDLP);
        
\end{circuitikz}

}
	\caption
	{%
	    Simplified representation of the power hardware of a NIC.
	}
	\label{fig:nic:hardware}
\end{figure}
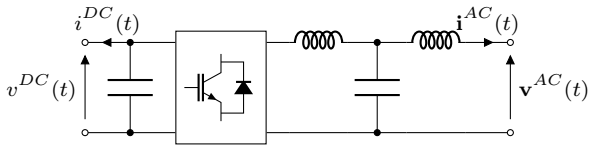

In this paper, the generic power hardware is modeled by the structure in \cref{fig:nic:hardware}.
It consists of an $LCL$ filter on the \AC side and a \DC-link capacitor on the \DC side.
Note that, as compared to the \CIDER in \cite{jrn:2022:becker}, a \NIC has an additionally interface to the \DC subsystem. 
Hence, the terminals of a \NIC are characterized by the \AC current $i^{\AC}$ and voltage $v^{\AC}$, plus the \DC current $i^{\DC}$ and voltage $v^{\DC}$.
The development of the time-domain state-space model of the power hardware including the \DC-link capacitor is explained in detail in Section~IV.A~of~\cite{jrn:2022:becker}.

Two different types of control laws are considered for the \NIC[s]: regulation of either i) the \DC-voltage and reactive power or ii) the active and reactive power injected on the \AC side.
In this paper, these two types of resources are referred to as $V_{\DC}/Q$-controlled~\NIC and $P/Q$-controlled~\NIC, respectively.
The time-domain state-space model of the control software of a $V_{\DC}/Q$-controlled~\NIC is shown in detail in Section~IV.B of \cite{jrn:2022:becker}.
The control software of a $P/Q$-controlled~\NIC is identical to the one of the grid-following \CIDER proposed in Section~III.C~of~\cite{jrn:2020:kettner-becker:HPF-2}.
Further details on the derivations can be found in the respective references.

\section{Resource Representation for \HPF Studies}
\label{sec:Resources}

From the time-domain state-space models of the resources, one can derive the harmonic-domain grid response.
For this purpose, the state-space models of power hardware and control software are first transformed to the harmonic domain by means of Fourier theor leading to Toeplitz matrices.
Then, they are combined to form the closed-loop model of the resource.
From this closed-loop model, the grid response of the resource is derived in the form of its harmonic transfer function.
Further details about this procedure can be found in Section~IV.B~of~\cite{jrn:2020:kettner-becker:HPF-1}.


\subsection{Types of Resources and Partition of the Nodes} %

The specific grid response of a resource describes its behaviour as seen from the point of connection to the grid.
A \emph{grid-forming} resource controls the magnitude and frequency of the grid voltage at its point of connection.
On the contrary, a \emph{grid-following} resource controls the injected current with a specific phase displacement w.r.t. the fundamental component of the grid voltage at its point of connection.
In line with these definitions, the set of all nodes $\nodes^{j}$ of a subsystem $j$ is partitioned into the disjoint sets $\formers^{j}$ and $\followers^{j}$, where grid-forming resources $s\in\formers^{j}$ and grid-following resources $r\in\followers^{j}$ are connected, respectively:
\begin{equation}
    \nodes^{j}=\formers^{j}\cup\followers^{j},~\formers^{j}\cap\followers^{j}=\emptyset
\end{equation}

In hybrid \AC/\DC grids, which consist of multiple subsystems, resources can be of single- or two-port type.
Single-port devices possess a single input/output terminal associated with one specific subsystem.
For instance, \CIDER[s] and impedance loads previously employed in \cite{jrn:2020:kettner-becker:HPF-2,jrn:2022:becker} are single-port devices.
Two-port devices are interconnections between two different subsystems.
For example, the \NIC[s] introduced in \cref{sec:NICs} have two ports (i.e., one each on the \AC and \DC side).
Following this reasoning, for each subsystem the sets $\followers^{j}$ and $\formers^{j}$ are further subdivided into two disjoint sets:
\begin{alignat}{2}
    \followers^{j}&=\followers^{j}_{1}\cup\followers^{j}_{2},~\followers^{j}_{1}\cap\followers^{j}_{2}&&=\emptyset \\
    \formers^{j}&=\formers^{j}_{1}\cup\formers^{j}_{2},\quad\formers^{j}_{1}\cap\formers^{j}_{2}&&=\emptyset
\end{alignat}
where $\followers^{j}_{1}$ and $\formers^{j}_{1}$ consist of all grid-forming and grid-following single-port resources (e.g., \CIDER[s]), respectively and $\followers^{j}_{2}$ and $\formers^{j}_{2}$ represent the nodes where the two-port resources (i.e., \NIC[s]) are connected.
The node partition of such a generic subsystem $j$ is shown in \cref{fig:nodes:j}.
In \cref{fig:nodes:hybrid} the example of a hybrid \AC/\DC grid consisting of one \AC and one \DC subsystem is shown.
In this example the two-port devices are purely composed of \NIC[s] that connect $\followers^{\AC}_2$ and $\formers^{\DC}_2$
Hence, the sets $\formers^{\AC}_2$ and $\followers^{\DC}_2$ are empty.
\begin{figure}[t]
    \centering
    \subfloat[]
    {%
        \centering
        {

\tikzstyle{set}=[circle, draw=black, minimum width=1.8cm, inner sep=0pt]

\scriptsize

\begin{tikzpicture}

	\def\dx{0.85}
	\def\dy{0.85}
	
	
	
	\node[set] (AC) at (-1.25*\dx,0){};
	\node at ($(AC.north)+(0,0.3*\dy)$) {Subsystem $j$};
    
    \draw[-] ($(AC.north)+(0,0.0*\dy)$)
    to ($(AC.south)+(0,-0.0*\dy)$);
        
    \node at ($(AC)+(-0.5*\dx,+0.4*\dy)$) {$\formers^{j}_{1}$};
    \node at ($(AC)+(-0.5*\dx,-0.4*\dy)$) {$\formers^{j}_{2}$};
    \node at ($(AC)+(+0.5*\dx,+0.4*\dy)$) {$\followers^{j}_{1}$};
    \node at ($(AC)+(+0.5*\dx,-0.4*\dy)$) {$\followers^{j}_{2}$};

    \draw[-,dashed] ($(AC.west)+(0,0.0*\dy)$)
    to ($(AC.east)+(0,-0.0*\dy)$);

    \draw[-,dotted,thick,white] ($(AC)+(-0.1*\dx,0.1*\dy)$) 
    to ($(AC.south)+(-0.1*\dx,-0.1*\dy)$)
    to ($(AC)+(-0.1*\dx,0.1*\dy)$);
    
\end{tikzpicture}

}
        \label{fig:nodes:j}
    }
    \subfloat[]
    {%
        \centering
        {

\tikzstyle{set}=[circle, draw=black, minimum width=1.8cm, inner sep=0pt]

\scriptsize

\begin{tikzpicture}

	\def\dx{0.85}
	\def\dy{0.85}
	
	
	\node[set] (AC) at (-1.25*\dx,0){};
	\node at ($(AC.north)+(0,0.3*\dy)$) {\AC Subsystem};
    
    \draw[-] ($(AC.north)+(0,0.0*\dy)$)
    to ($(AC.south)+(0,-0.0*\dy)$);
        
    \node at ($(AC)+(-0.5*\dx,+0.0*\dy)$) {$\formers^{\AC}_{1}$};
    \node at ($(AC)+(+0.5*\dx,+0.4*\dy)$) {$\followers^{\AC}_{1}$};
    \node at ($(AC)+(+0.5*\dx,-0.4*\dy)$) {$\followers^{\AC}_{2}$};

    \draw[-,dashed] ($(AC)+(0,0.0*\dy)$)
    to ($(AC.east)+(0,-0.0*\dy)$);
    
	\node[set] (DC) at (1.1*\dx,0){};
	\node at ($(DC.north)+(0,0.3*\dy)$) {\DC Subsystem};
    
    \draw[-] ($(DC.north)+(0,0.0*\dy)$)
    to ($(DC.south)+(0,-0.0*\dy)$);
        
    \node at ($(DC)+(-0.45*\dx,+0.4*\dy)$) {$\formers^{\DC}_{1}$};
    \node at ($(DC)+(-0.45*\dx,-0.4*\dy)$) {$\formers^{\DC}_{2}$};
    \node at ($(DC)+(+0.55*\dx,-0.0*\dy)$) {$\followers^{\DC}_{1}$};

    \draw[-,dashed] ($(DC)$) to ($(DC.west)$);

    \draw[-,dotted,thick,green] ($(AC)+(-0.1*\dx,0.1*\dy)$) 
    to ($(DC)+(0.1*\dx,0.1*\dy)$)
    to ($(DC.south)+(0.1*\dx,-0.1*\dy)$)
    to ($(AC.south)+(-0.1*\dx,-0.1*\dy)$)
    to ($(AC)+(-0.1*\dx,0.1*\dy)$);
    
	\node[green] at ($0.5*(DC.south)+0.5*(AC.south)+(0,+0.2*\dy)$) {\NIC[s]};
 
    
\end{tikzpicture}

}
        \label{fig:nodes:hybrid}
    }
	\caption
	{%
        Partition of the nodes for a generic subsystem $j$ (\cref{fig:nodes:j}) and for a hybrid \AC/\DC grid with \NIC[s] connecting $\followers^{\AC}_2$ and $\formers^{\DC}_2$ (\cref{fig:nodes:hybrid}).
	}
	\label{fig:nodes}
\end{figure}


\subsection{Grid Response of Single-Port Resources} %

As previously stated, the grid response of a single-port resource defines either the voltage in function of the current or vice versa.
As it was introduced in Section~II.C~of~\cite{jrn:2022:becker}, the grid responses of grid-forming and grid-following \CIDER[s] can be expressed as follows:
\begin{alignat}{2}
        s\in\formers^{j}_1:\quad
    &   \Hat{\mathbf{V}}^{j}_{s}
    &&=    \Hat{\mathbf{Y}}_{s}(\Hat{\mathbf{I}}^{j}_{s},\Hat{\mathbf{W}}_{\spt,s},\hat{\mathbf{Y}}_{o,s})
    \label{eq:cider:form}\\
        r\in\followers^{j}_1:\quad
    &   \Hat{\mathbf{I}}^{j}_{r}
    &&=    \Hat{\mathbf{Y}}_{r}(\Hat{\mathbf{V}}^{j}_{r},\Hat{\mathbf{W}}_{\spt,r},\hat{\mathbf{Y}}_{o,r})
    \label{eq:cider:follow}
\end{alignat}
where $\Hat{\mathbf{V}}^{j}_{s}$ and $\Hat{\mathbf{I}}^{j}_{s}$ are the column vectors of the Fourier coefficients of the terminal voltages $\mathbf{v}^{j}_{s}(t)$ and currents $\mathbf{i}^{j}_{s}(t)$, respectively, of the grid-forming resources.
Similarly, $\Hat{\mathbf{V}}^{j}_{r}$ and $\Hat{\mathbf{I}}^{j}_{r}$ represent the column vectors of the terminal voltages and currents of the grid-following resources.
The inputs $\Hat{\mathbf{W}}_{\spt,s}$ and $\hat{\mathbf{Y}}_{o,s}$ represent the setpoints and the operating points, respectively, of the \CIDER[s].
The latter is needed in case a linearization was performed in the derivation of the \CIDER model, as presented in Section~II.C~of~\cite{jrn:2022:becker}.
It is worth noting that this representation of the grid response is generic (e.g., it also applies to passive loads) and that it is valid irrespective of the type of grid it is connected to (i.e., \AC or \DC).
For further details on the derivation of the grid response, please refer to Section~II.C~of~\cite{jrn:2022:becker}.


\subsection{Grid Response of Two-Port Resources}

The grid response of a two-port device is described by two pairs of electrical quantities (i.e., voltages and currents at both ports).
It is important to note that a \NIC cannot exhibit grid-forming and grid-following behaviour arbitrarily at each port as the corresponding circuit equations cannot be overdetermined.
To be more precise, one cannot simultaneously control the voltages or currents at both ports.
By consequence, if one port exhibits a grid-forming behaviour, the other must exhibit a grid-following behaviour.

Typically, \NIC[s] perform grid-forming control on the \DC side and grid-following control on the \AC side%
\footnote{
    In theory, it is possible to perform grid-forming control on the \AC side and grid-following control on the \DC side.
    In practice, this configuration is not employed to the best of the authors' knowledge.
}.
Thus, a \NIC can be seen as a branch $m=(r,s)$ between two nodes $r\in\followers^{\AC}_2$ and $s\in\formers^{\DC}_2$ (i.e., $\mathcal{M}\subseteq\followers^{\AC}_2\times\formers^{\DC}_2$).
The grid response of a \NIC $m=(r,s)\in\mathcal{M}$ is described by
\begin{alignat}{2}
        [\Hat{\mathbf{I}}^{\AC}_{r},\Hat{\mathbf{V}}^{\DC}_{s}]
    &=    \Hat{\mathbf{Y}}_{m}(\Hat{\mathbf{V}}^{\AC}_{r},\Hat{\mathbf{I}}^{\DC}_{s},\Hat{\mathbf{W}}_{\spt,m},\hat{\mathbf{Y}}_{o,m})
    \label{eq:nic}
\end{alignat}
As for the case of \CIDER[s], $\Hat{\mathbf{W}}_{\spt,m}$ and $\hat{\mathbf{Y}}_{o,m}$ denote the setpoint and operating point, respectively.

\section{\HPF Algorithm for Hybrid \AC/\DC Grids}
\label{sec:Algorithm}

The \HPF framework proposed in \cite{jrn:2020:kettner-becker:HPF-1} describes \AC power systems by two sets of nodal equations.
Namely, the nodal quantities are expressed from the point of view of the grid and the resources, respectively.
Recall from \cref{sec:Resources} the partition of the nodes of each subsystem into the sets $\formers^{j}$ and $\followers^{j}$.
The unknowns of the \HPF problem are the nodal injected currents at the nodes $\formers^{j}$ and the nodal phase-to-ground voltages at the nodes $\followers^{j}$ (i.e., the quantities that are not regulated by the respective type of resource).
The nodal equations of a hybrid \AC/\DC grid are obtained as the combination of the nodal equations of all subsystems.

\subsection{Mismatch Equations}

The nodal equations for each subsystem $j$ seen from the grid are formulated using hybrid parameters:
\begin{align}
        \Hat{\mathbf{V}}_{\formers}^{j}
    &=      \Hat{\mathbf{H}}^{j}_{\formers\times\formers}\Hat{\mathbf{I}}^{j}_{\formers}
        +   \Hat{\mathbf{H}}^{j}_{\formers\times\followers}\Hat{\mathbf{V}}^{j}_{\followers}
    \label{eq:grid:form}
    \\
        \Hat{\mathbf{I}}_{\followers}^{j}
    &=      \Hat{\mathbf{H}}^{j}_{\followers\times\formers}\Hat{\mathbf{I}}^{j}_{\formers}
        +   \Hat{\mathbf{H}}^{j}_{\followers\times\followers}\Hat{\mathbf{V}}^{j}_{\followers}
    \label{eq:grid:follow}
\end{align}
where $\Hat{\mathbf{H}}^{j}_{\formers\times\formers}$, $\Hat{\mathbf{H}}^{j}_{\formers\times\followers}$, $\Hat{\mathbf{H}}^{j}_{\followers\times\formers}$ and $\Hat{\mathbf{H}}^{j}_{\followers\times\followers}$ are the blocks of the hybrid matrix $\Hat{\mathbf{H}}^{j}$ associated with $\formers^{j}$ and $\followers^{j}$.
$\Hat{\mathbf{I}}^{j}_{\formers}$ and $\Hat{\mathbf{V}}^{j}_{\followers}$ are the column vectors of all nodal injected currents at the nodes $\formers^{j}$ and all nodal phase-to-ground voltages at the nodes $\followers^{j}$, respectively.
For further details, please see Section~III~of~\cite{jrn:2020:kettner-becker:HPF-1}.

The \HPF problem is given by the mismatch equations between \eqref{eq:grid:form}--\eqref{eq:grid:follow} on the one hand and the grid responses of the resources \eqref{eq:cider:form}--\eqref{eq:cider:follow} and \eqref{eq:nic} on the other hand.
At the equilibrium, these mismatches must be zero.
Thus, at the nodes where single-port resources are connected:
\begin{align}
        \Delta\Hat{\mathbf{V}}^{j}_{\formers_1}
        (\Hat{\mathbf{I}}^{j}_{\formers},\Hat{\mathbf{V}}^{j}_{\followers},\Hat{\mathbf{W}}^{j}_{\sigma,\formers_1},\Hat{\mathbf{Y}}^{j}_{o,\formers_1})
    &=  \mathbf{0}
    \label{eq:residual:form1}\\
        \Delta\Hat{\mathbf{I}}^{j}_{\followers_1}
        (\Hat{\mathbf{I}}^{j}_{\formers},\Hat{\mathbf{V}}^{j}_{\followers},\Hat{\mathbf{W}}^{j}_{\sigma,\followers_1},\Hat{\mathbf{Y}}^{j}_{o,\followers_1})
    &=  \mathbf{0}
    \label{eq:residual:follow1}
\end{align}
Notably, this formulation corresponds to the one introduced in Section~III~of~\cite{jrn:2022:becker}.
At the nodes where two-port resources are connected, the argument of the mismatch equations has to be expanded to account for the quantitiy of the second terminal of the \NIC[s].
Thus,
\begin{align}
        \Delta\Hat{\mathbf{V}}^{\DC}_{\formers_2}
        (\Hat{\mathbf{I}}^{\DC}_{\formers},\Hat{\mathbf{V}}^{\DC}_{\followers},\Hat{\mathbf{V}}^{\AC}_{\followers_2},\Hat{\mathbf{W}}^{\DC}_{\sigma,\formers_2},\Hat{\mathbf{Y}}^{\DC}_{o,\formers_2})
    &=  \mathbf{0}
    \label{eq:residual:form2}\\
        \Delta\Hat{\mathbf{I}}^{\AC}_{\followers_2}
        (\Hat{\mathbf{I}}^{\AC}_{\formers},\Hat{\mathbf{V}}^{\AC}_{\followers},\Hat{\mathbf{I}}^{\DC}_{\formers_2},\Hat{\mathbf{W}}^{\AC}_{\sigma,\followers_2},\Hat{\mathbf{Y}}^{\AC}_{o,\followers_2})
    &=  \mathbf{0}
    \label{eq:residual:follow2}
\end{align}
where $\Hat{\mathbf{V}}^{\AC}_{\followers_2}$ and $\Hat{\mathbf{I}}^{\DC}_{\formers_2}$ represent the coupling between the two subsystems.
This system of a equations can be solved using the Newton-Raphson algorithm.
To this end, the Jacobian matrix of the equations is required.


\subsection{Jacobian Matrix}

The Jacobian matrix $\Hat{\mathbf{J}}$ required for the Newton-Raphson algorithm is derived as the difference between the Jacobian matrices of the equations related to the resources $\Hat{\mathbf{J}}^{RSC}$ and the grid $\Hat{\mathbf{J}}^{GRD}$, respectively.
That is,
\begin{align}
        \Hat{\mathbf{J}}
    &=  \Hat{\mathbf{J}}^{RSC}-\Hat{\mathbf{J}}^{GRD}
\end{align}

$\Hat{\mathbf{J}}^{GRD}$ is composed of the partial derivatives of the grid equations in \eqref{eq:grid:form}--\eqref{eq:grid:follow} w.r.t. $\Hat{\mathbf{I}}^{j}_{\formers}$ and $\Hat{\mathbf{V}}^{j}_{\followers}$, respectively.
More precisely, it is composed of the hybrid parameters of the \AC and \DC subsystem:
{
\renewcommand\arraystretch{1.3}
\begin{align}
        \Hat{\mathbf{J}}^{GRD}
	&=		\left[
			\begin{array}{ll|ll}
				    \Hat{\mathbf{H}}^{\AC}_{\formers\times\formers}
				&   \Hat{\mathbf{H}}^{\AC}_{\formers\times\followers}
				&   \mathbf{0}
				&   \mathbf{0}\\
				    \Hat{\mathbf{H}}^{\AC}_{\followers\times\formers}
				&   \Hat{\mathbf{H}}^{\AC}_{\followers\times\followers}
				&   \mathbf{0}
				&   \mathbf{0}\\ \hline
				      \mathbf{0}
				&   \mathbf{0}
				&   \Hat{\mathbf{H}}^{\DC}_{\formers\times\formers}
				&   \Hat{\mathbf{H}}^{\DC}_{\formers\times\followers}\\
				    \mathbf{0}
				&   \mathbf{0}
				&   \Hat{\mathbf{H}}^{\DC}_{\followers\times\formers}
				&   \Hat{\mathbf{H}}^{\DC}_{\followers\times\followers}
			\end{array}
			\right]
    \label{eq:jacobian:hybrid:grid}
\end{align}
}
Notably, this matrix has a block-diagonal structure, and exhibits no coupling between \AC and \DC quantities.

$\Hat{\mathbf{J}}^{RSC}$ is composed of the partial derivatives of the grid responses in \eqref{eq:cider:form}--\eqref{eq:cider:follow} and \eqref{eq:nic} w.r.t. $\Hat{\mathbf{I}}^{j}_{\formers}$ and $\Hat{\mathbf{V}}^{j}_{\followers}$, respectively.
The following notation is introduced in the following for the sake of conciseness: 
\begin{align}
    \partial^{j}_{\formers_k}{} = \derivative{}{\Hat{\mathbf{I}}^{j}_{\formers_k}},\quad
    \partial^{j}_{\followers_k}{} = \derivative{}{\Hat{\mathbf{V}}^{j}_{\followers_k}} 
\end{align}
$\Hat{\mathbf{J}}^{RSC}$ is depicted in \eqref{eq:jacobian:hybrid:res}.
At nodes with single-port resources, the corresponding blocks of the Jacobian matrix exhibit a diagonal structure, since a resource influences only the quantity at its point of connection.
These terms associated with the \AC and \DC single-port resources are highlighted in blue and red, respectively, in \eqref{eq:jacobian:hybrid:res}.
At nodes with two-port resources (i.e., where \NIC[s] are connected), off-diagonal terms appear in the Jacobian matrix.
They describe the coupling between the \AC and \DC port of the \NIC[s], respectively, as well as between the \AC and \DC subsystems.
These terms are highlighted in green in \eqref{eq:jacobian:hybrid:res}.
\begin{figure*}[h] 
\begin{equation}  
        \Hat{\mathbf{J}}^{RSC}   
    =       \begin{tikzpicture}[baseline] 

     
    \def\dx{8pt}
    \def\dy{6pt}
    
    \matrix [matrix of math nodes,left delimiter=[,right delimiter={]}] (m) 
    {
            |(11)|  \textcolor{blue}{\partial^{\AC}_{\formers_1}{\Hat{\mathbf{V}}^{\AC}_{\formers_1}}}
        &   |(12)|  \mathbf{0}
        &   |(13)|  \mathbf{0}
        &   |(14)|  \mathbf{0}
        &   |(15)|  \mathbf{0}
        &   |(16)|  \mathbf{0}\\
            |(21)|  \mathbf{0}
        &   \textcolor{blue}{\partial^{\AC}_{\followers_1}{\Hat{\mathbf{I}}^{\AC}_{\followers_1}}}
        &   \mathbf{0}
        &   \mathbf{0}
        &   \mathbf{0}
        &   |(26)|  \mathbf{0}\\
            |(31)|  \mathbf{0}
        &   \mathbf{0}
        &   \partial^{\AC}_{\followers_2}{\Hat{\mathbf{I}}^{\AC}_{\followers_2}}
        &   \mathbf{0}
        &   \textcolor{green}{\partial^{\DC}_{\formers_2}{\Hat{\mathbf{I}}^{\AC}_{\followers_2}}}
        &   |(36)|  \mathbf{0}\\
            |(41)|  \mathbf{0}
        &   \mathbf{0}
        &   \mathbf{0}
        &   \textcolor{red}{\partial^{\DC}_{\formers_1}{\Hat{\mathbf{V}}^{\DC}_{\formers_1}}}
        &   \mathbf{0}
        &   |(46)|  \mathbf{0}\\ 
            |(51)|  \mathbf{0}
        &   \mathbf{0}
        &   \textcolor{green}{\partial^{\AC}_{\followers_2}{\Hat{\mathbf{V}}^{\DC}_{\formers_2}}}
        &   \mathbf{0}
        &   \partial^{\DC}_{\followers_2}{\Hat{\mathbf{V}}^{\DC}_{\formers_2}}
        &   |(56)|  \mathbf{0}\\ 
            |(61)|  \mathbf{0}
        &   |(62)|  \mathbf{0}
        &   |(63)|  \mathbf{0}
        &   |(64)|  \mathbf{0}
        &   |(65)|  \mathbf{0}
        &   |(66)|  \textcolor{red}{\partial^{\DC}_{\followers_1}{\Hat{\mathbf{I}}^{\DC}_{\followers_1}}}\\
    };
        
    \node[above=\dy of 11.center] (top-1) {}; 
    \node[above=\dy of 12.center] (top-2) {}; 
    \node[above=\dy of 13.center] (top-3) {};
    \node[above=\dy of 14.center] (top-4) {}; 
    \node[above=\dy of 15.center] (top-5) {}; 
    \node[above=\dy of 16.center] (top-6) {};
    
    \node[below=\dy of 61.center] (bottom-1) {}; 
    \node[below=\dy of 62.center] (bottom-2) {}; 
    \node[below=\dy of 63.center] (bottom-3) {}; 
    \node[below=\dy of 64.center] (bottom-4) {}; 
    \node[below=\dy of 65.center] (bottom-5) {}; 
    \node[below=\dy of 66.center] (bottom-6) {}; 
    
    \node[left=\dx of 11.center] (left-1) {}; 
    \node[left=\dx of 21.center] (left-2) {}; 
    \node[left=\dx of 31.center] (left-3) {}; 
    \node[left=\dx of 41.center] (left-4) {};
    \node[left=\dx of 51.center] (left-5) {}; 
    \node[left=\dx of 61.center] (left-6) {};
    
    \node[right=\dx of 16.center] (right-1) {}; 
    \node[right=\dx of 26.center] (right-2) {}; 
    \node[right=\dx of 36.center] (right-3) {}; 
    \node[right=\dx of 46.center] (right-4) {};
    \node[right=\dx of 56.center] (right-5) {}; 
    \node[right=\dx of 66.center] (right-6) {};
    
    \draw[-] ($0.5*(top-3)+0.5*(top-4)$) -- ($0.5*(bottom-3)+0.5*(bottom-4)$);
    \draw[-] ($0.5*(left-3)+0.5*(left-4)$) -- ($0.5*(right-3)+0.5*(right-4)$);
    
    \draw[dashed] ($0.5*(top-1)+0.5*(top-2)$) -- ($0.5*(bottom-1)+0.5*(bottom-2)$);
    \draw[dashed] ($0.5*(left-1)+0.5*(left-2)$) -- ($0.5*(right-1)+0.5*(right-2)$);
    
    \draw[dashed] ($0.5*(top-5)+0.5*(top-6)$) -- ($0.5*(bottom-5)+0.5*(bottom-6)$);
    \draw[dashed] ($0.5*(left-5)+0.5*(left-6)$) -- ($0.5*(right-5)+0.5*(right-6)$);

    
     
    \draw[thick,decorate,decoration=brace] ($(top-1.west)+(-0.2,0.4)$) -- ($(top-1.east)+(0.2,0.4)$)
    node[midway,above=1mm] {$\formers^{\AC}$};   
    \draw[thick,decorate,decoration=brace] ($(top-2.west)+(-0.2,0.4)$) -- ($(top-3.east)+(0.2,0.4)$)
    node[midway,above=1mm] {$\followers^{\AC}$};
    
    \draw[thick,decorate,decoration=brace] ($(top-4.west)+(-0.2,0.4)$) -- ($(top-5.east)+(0.2,0.4)$)
    node[midway,above=1mm] {$\formers^{\DC}$};
    \draw[thick,decorate,decoration=brace] ($(top-6.west)+(-0.2,0.4)$) -- ($(top-6.east)+(0.2,0.4)$)
    node[midway,above=1mm] {$\followers^{\DC}$};
    
    

    \draw[thick,decorate,decoration=brace] ($(right-1.north)+(0.9,0.1)$) -- ($(right-1.south)+(0.9,-0.1)$)
    node[midway,right=1mm] {$\formers^{\AC}$};
    \draw[thick,decorate,decoration=brace] ($(right-2.north)+(0.9,0.1)$) -- ($(right-3.south)+(0.9,-0.1)$)
    node[midway,right=1mm] {$\followers^{\AC}$};
    
    \draw[thick,decorate,decoration=brace] ($(right-4.north)+(0.9,0.1)$) -- ($(right-5.south)+(0.9,-0.1)$)
    node[midway,right=1mm] {$\formers^{\DC}$};
    \draw[thick,decorate,decoration=brace] ($(right-6.north)+(0.9,0.1)$) -- ($(right-6.south)+(0.9,-0.1)$)
    node[midway,right=1mm] {$\followers^{\DC}$};
\end{tikzpicture} 
    \label{eq:jacobian:hybrid:res}
\end{equation}
\end{figure*}

Details on the derivation of the partial derivatives in \eqref{eq:jacobian:hybrid:grid} and \eqref{eq:jacobian:hybrid:res} can be found in Section~V.B~of~\cite{jrn:2020:kettner-becker:HPF-1}.

Using the aforestated expression for the mismatch equations and the Jacobian matrix, the \HPF problem can be solved by means of the Newton-Raphson algorithm as described in Section~III~of~\cite{jrn:2022:becker}.

\section{Validation}
\label{sec:val-sys}


\subsection{Methodology and Key Performance Indicators}
\label{sec:val-sys:method}

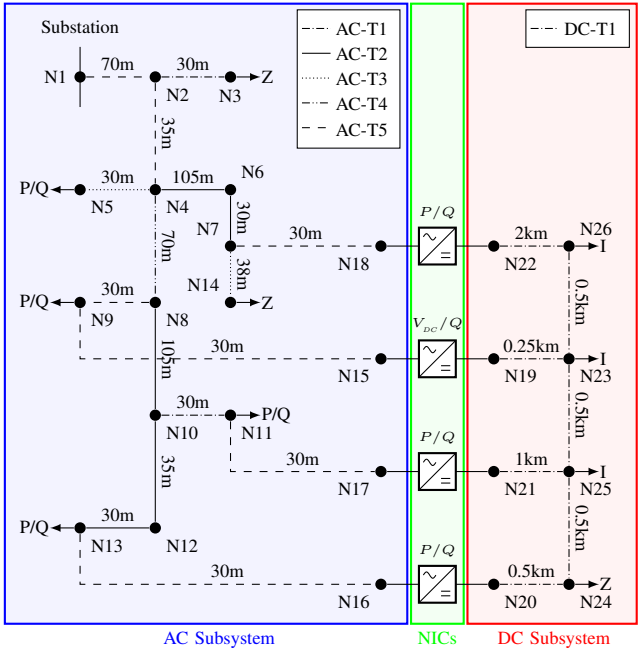
\begin{figure}[t]
	\centering
    {
\tikzstyle{bus}=[circle,fill=black,minimum size=1.5mm,inner sep=0mm]
\tikzstyle{SILA}=[densely dashdotted]
\tikzstyle{SILB}=[]
\tikzstyle{SILC}=[densely dotted]
\tikzstyle{SILD}=[densely dashdotdotted]
\tikzstyle{SILE}=[dashed]
\tikzstyle{load}=[-latex]

\begin{circuitikz}
	\scriptsize
	
	\def\BlockSize{1.0}	
	\def\X{3.8}	
	\def\Y{3.6}	
	\def\dlX{1.0}
	\def\dlY{1.5}
	\def\dload{0.4}


    \draw[thick,blue,fill=blue, fill opacity=0.05] (-1*\dlX,0.65*\dlY) rectangle ++(5.35*\dlX,-5.5*\dlY) node[below,shift={(-2.5*\dlX,0)},opacity=1] {AC Subsystem};
    \draw[thick,red,fill=red, fill opacity=0.05]  (5.15*\dlX,0.65*\dlY) rectangle ++(2.25*\dlX,-5.5*\dlY) node[below,shift={(-1.1*\dlX,0)},opacity=1] {DC Subsystem}; 
    \draw[thick,green,fill=green, fill opacity=0.05]  (4.4*\dlX,0.65*\dlY) rectangle ++(0.7*\dlX,-5.5*\dlY) node[below,shift={(-0.35*\dlX,0)},opacity=1] {NICs}; 
    
    
	\node[bus,label={left:N1}] (N1) at (0,0) {};
	\node[bus,label={below right:N2}] (N2) at ($(N1) + (\dlX,0)$) {};
	\node[bus,label={below:N3}] (N3) at ($(N2) + (\dlX,0)$) {};
 
	\node[bus,label={below right:N4}] (N4) at ($(N2) + (0,-\dlY)$) {};
	\node[bus,label={below right:N5}] (N5) at ($(N4) + (-\dlX,0)$) {};
	\node[bus,label={above right:N6}] (N6) at ($(N4) + (\dlX,0)$) {};
	\node[bus,label={above left:N7}] (N7) at ($(N6) + (0*\dlX,-0.5*\dlY)$) {};
	\node[bus,label={above left:N14}] (N14) at ($(N7) + (0*\dlX,-0.5*\dlY)$) {};
 
	\node[bus,label={below right:N8}] (N8) at ($(N4) + (0,-\dlY)$) {};
	\node[bus,label={below right:N9}] (N9) at ($(N8) + (-\dlX,0)$) {};
 
	\node[bus,label={below right:N10}] (N10) at ($(N8) + (0,-\dlY)$) {};
	\node[bus,label={below right:N11}] (N11) at ($(N10) + (\dlX,0)$) {};
 
	\node[bus,label={below right:N12}] (N12) at ($(N10) + (0,-\dlY)$) {};
	\node[bus,label={below right:N13}] (N13) at ($(N12) + (-\dlX,0)$) {};

	\node[bus,label={below left:N18}] (N18) at (4*\dlX,-1.5*\dlY) {};
	\node[bus,label={below left:N15}] (N15) at (4*\dlX,-2.5*\dlY) {};
	\node[bus,label={below left:N17}] (N17) at (4*\dlX,-3.5*\dlY) {};
	\node[bus,label={below left:N16}] (N16) at (4*\dlX,-4.5*\dlY) {};
    
	\node[bus,label={below right:N22}] (N22) at (5.5*\dlX,-1.5*\dlY) {};
	\node[bus,label={below right:N19}] (N19) at (5.5*\dlX,-2.5*\dlY) {};
	\node[bus,label={below right:N21}] (N21) at (5.5*\dlX,-3.5*\dlY) {};
	\node[bus,label={below right:N20}] (N20) at (5.5*\dlX,-4.5*\dlY) {};
 
	\node[bus,label={above right:N26}] (N26) at ($(N22) + (\dlX,0)$) {};
	\node[bus,label={below right:N23}] (N23) at ($(N19) + (\dlX,0)$) {};
	\node[bus,label={below right:N25}] (N25) at ($(N21) + (\dlX,0)$) {};
	\node[bus,label={below right:N24}] (N24) at ($(N20) + (\dlX,0)$) {};

	
	\draw[SILE] (N1) to node[midway,above]{70m} (N2) {};
	\draw[SILA] (N2) to node[midway,above]{30m} (N3) {};
 
	\draw[SILE] (N2) to node[sloped,anchor=center,above]{35m} (N4) {};
	\draw[SILC] (N4) to node[midway,above]{30m} (N5) {};
	\draw[SILB] (N4) to node[midway,above]{105m} (N6) {};
	\draw[SILB] (N6) to node[sloped,anchor=center,above]{30m} (N7) {};
	\draw[SILC] (N7) to node[sloped,anchor=center,above]{38m} (N14) {};
 
	\draw[SILD] (N4) to node[sloped,anchor=center,above]{70m} (N8) {};
	\draw[SILE] (N8) to node[midway,above]{30m} (N9) {};
	
	\draw[SILB] (N8) to node[sloped,anchor=center,above]{105m} (N10) {};
	\draw[SILA] (N10) to node[midway,above]{30m} (N11) {};
	
	\draw[SILB] (N10) to node[sloped,anchor=center,above]{35m} (N12) {};
	\draw[SILB] (N12) to node[midway,above]{30m} (N13) {};
 
 
	\draw[SILE]  (N7) 
        to node[midway,above]{30m} (N18) {};
	\draw[SILE] (N9) 
        to ($(N9)-(0,0.5*\dlY)$)
        to node[midway,above]{30m} (N15) {};
	\draw[SILE] (N11) 
        to ($(N11)-(0,0.5*\dlY)$)
        to node[midway,above]{30m} (N17) {};
	\draw[SILE] (N13) 
        to ($(N13)-(0,0.5*\dlY)$)
        to node[midway,above]{30m} (N16) {};

    
	\draw[SILA] (N19) to node[midway,above]{0.25km} (N23) {};
	\draw[SILA] (N20) to node[midway,above]{0.5km} (N24) {};
	\draw[SILA] (N21) to node[midway,above]{1km} (N25) {};
	\draw[SILA] (N22) to node[midway,above]{2km} (N26) {};
 
	\draw[SILA] (N23) to node[sloped,anchor=center,above]{0.5km} (N25) {};
	\draw[SILA] (N26) to node[sloped,anchor=center,above]{0.5km} (N23) {};
	\draw[SILA] (N25) to node[sloped,anchor=center,above]{0.5km} (N24) {};

 
 
    \node [sacdcshape,scale=0.5,fill=white,label={above,font=\tiny:$V_{\scaleto{\DC}{1.75pt}}/Q$}](ACDC1) at ($0.5*(N15)+0.5*(N19)$) {};
	\draw[short] (ACDC1.west) to (N15) {};
	\draw[short] (ACDC1.east) to (N19) {};
    \node [sacdcshape,scale=0.5,fill=white,label={above,font=\tiny:$P/Q$}](ACDC2) at ($0.5*(N16)+0.5*(N20)$) {};
	\draw[short] (ACDC2.west) to (N16) {};
	\draw[short] (ACDC2.east) to (N20) {};
    \node [sacdcshape,scale=0.5,fill=white,label={above,font=\tiny:$P/Q$}](ACDC3) at ($0.5*(N17)+0.5*(N21)$) {};
	\draw[short] (ACDC3.west) to (N17) {};
	\draw[short] (ACDC3.east) to (N21) {};
    \node [sacdcshape,scale=0.5,fill=white,label={above,font=\tiny:$P/Q$}](ACDC4) at ($0.5*(N18)+0.5*(N22)$) {};
	\draw[short] (ACDC4.west) to (N18) {};
	\draw[short] (ACDC4.east) to (N22) {};
    
    
	\draw[load] (N5) to node[left,align=right]{~P/Q}
	($(N5)-\dload*(1,0)$);
	\draw[load] (N9) to node[left,align=right]{~P/Q}
	($(N9)-\dload*(1,0)$);
	\draw[load] (N11) to node[right,align=left]{~P/Q}
	($(N11)+\dload*(1,0)$);
	\draw[load] (N13) to node[left,align=right]{~P/Q}
	($(N13)-\dload*(1,0)$);
	
	\draw[load] (N3) to node[right,align=left]{~Z}
	($(N3)+\dload*(1,0)$);
	\draw[load] (N14) to node[right,align=left]{~Z}
	($(N14)+\dload*(1,0)$);
 
	\draw[load] (N23) to node[right,align=left]{~I}
	($(N23)+\dload*(1,0)$);
	\draw[load] (N25) to node[right,align=left]{~I}
	($(N25)+\dload*(1,0)$);
	\draw[load] (N26) to node[right,align=left]{~I}
	($(N26)+\dload*(1,0)$);
	
	\draw[load] (N24) to node[right,align=left]{~Z}
	($(N24)+\dload*(1,0)$);

	\draw[-] ($(N1)+\dload*(0,-1)$) to ($(N1)+\dload*(0,1)$);
	\node[label={above:Substation}] (Substation) at ($(N1)+\dload*(0,1)$) {};

	
    \coordinate (LegAC) at (2.88*\dlX,0.6*\dlY);
    \matrix [draw,below right,fill=white] at (LegAC) {
        \node [SILA,label=right:~~AC-T1] {}; \\
        \node [SILB,label=right:~~AC-T2] {}; \\
        \node [SILC,label=right:~~AC-T3] {}; \\
        \node [SILD,label=right:~~AC-T4] {}; \\
        \node [SILE,label=right:~~AC-T5] {}; \\
    };
    
	\def\dl{-0.33}
	\def\do{-0.17*\dlY}
    
    \draw[SILA] ($(LegAC)+(0.07*\dlX,\do)$) to node[]{} ($(LegAC)+(0.45*\dlX,\do)$);
    \draw[SILB] ($(LegAC)+(0.07*\dlX,\do)+(0,\dl)$) to node[]{} ($(LegAC)+(0.45*\dlX,\do)+(0,\dl)$);
    \draw[SILC] ($(LegAC)+(0.07*\dlX,\do)+(0,2*\dl)$) to node[]{} ($(LegAC)+(0.45*\dlX,\do)+(0,2*\dl)$);
    \draw[SILD] ($(LegAC)+(0.07*\dlX,\do)+(0,3*\dl)$) to node[]{} ($(LegAC)+(0.45*\dlX,\do)+(0,3*\dl)$);
    \draw[SILE] ($(LegAC)+(0.07*\dlX,\do)+(0,4*\dl)$) to node[]{} ($(LegAC)+(0.45*\dlX,\do)+(0,4*\dl)$);

	
    \coordinate (LegDC) at (5.92*\dlX,0.6*\dlY);
    \matrix [draw,below right,fill=white] at (LegDC) {
        \node [SILA,label=right:~~DC-T1] {}; \\
    };
    
	\def\dl{-0.38}
	\def\do{-0.17*\dlY}
 
    \draw[SILA] ($(LegDC)+(0.07*\dlX,\do)$) to node[]{} ($(LegDC)+(0.45*\dlX,\do)$);

\end{circuitikz}
}
	\caption
	{%
	    Schematic diagram of the test system, which is based on the \CIGRE low-voltage benchmark microgrid \cite{Rep:2005:CIGRE} (blue box) and interfaced through NICs (green box) to the \DC subsystem (red box), parameters given in \cref{tab:nics:references}.
	    The resources are composed of constant impedance loads (Z), constant power loads (P/Q), and constant current sources (I), parameters given in \cref{tab:resources:references}.
	}
	\label{fig:grid:schematic}
\end{figure}

The \HPF algorithm for hybrid \AC/\DC grids is validated on an extension of the benchmark \AC microgrid proposed in \cite{Rep:2005:CIGRE}.
More specifically, the benchmark system is extended by a \DC grid following the example of \cite{jrn:2022:lambrichts}.
As depicted in \cref{fig:grid:schematic} the \AC and \DC subsystems are interfaced through \NIC[s] at the nodes N15-18 on the \AC side and N19-22 on the \DC side.
Their specifications are given in \cref{tab:nics:references}.
The \AC subsystem is composed of a feeding substation at node N1, seven grid-following \CIDER[s] at nodes N5, N9, N11 and N13 and two passive loads at nodes N3 and N14.
The \DC subsystem consists of three current sources at nodes N23 and N25-26, and a passive load at node N24.
The references of the resources are given in \cref{tab:resources:references}.
The feeding substation of the \AC subsystem is modelled as a \TE described by parameters depicted in \cref{tab:TE:parameters}.
The \TE injects harmonics with levels shown in \cref{tab:TE:harmonics} based on \cite{Std:BSI-EN-50160:2000}.
The line parameters of the \AC and \DC subsystems are given in \cref{tab:grid:parameters}.

The validation of the \HPF algorithm is performed through \emph{Time-Domain Simulations} (\TDS) in Simulink.
To this end, the system in \cref{fig:grid:schematic} is replicated using the models of the \CIDER[s] in \cite{jrn:2022:becker}.
The Matlab code of the \HPF method is updated to account for hybrid \AC/\DC grids.
For the \TDS, a \emph{Discrete Fourier Transform} (\DFT) over 5 periods of the fundamental frequency in steady state is performed.
All signals are normalized w.r.t. the base power $P_{b}=50\,\text{kW}$ and base voltage $V_{b}=230\,\text{V-\RMS}$.

\begin{table}[t]
    \centering
    \caption{Parameters of Network-Interfacing Converters.}
    \label{tab:nics:references}
	{

\renewcommand{\arraystretch}{1.2}
\begin{tabular}{cccccc}
	\hline
		\AC~Node
	&	\DC~Node
	&	P
	&	Q
	&	$V_{\DC}$
	&	Type
	\\
	\hline
		N15
	&	N19
	&	-
	&	\phantom{-}9.9~kVar
	&   900~V
	&   $V_{\DC}$/Q
	\\
		N16
	&	N20
	&	\phantom{-}30.0~kVar
	&	\phantom{-}9.9~kVar
	&   -
	&   P/Q
	\\
		N17
	&	N21
	&	-25.0~kVar
	&	8.2~kVar
	&   -
	&   P/Q
	\\
		N18
	&	N22
	&	\phantom{-}30.0~kVar
	&	\phantom{-}9.9~kVar
	&   -
	&   P/Q
	\\
	\hline
\end{tabular}
}
\end{table}

\begin{table}[t]
    \centering
    \caption{Parameters of the Grid-Following Resources and Loads.}
    \label{tab:resources:references}
	{

\renewcommand{\arraystretch}{1.2}
\begin{tabular}{cccc}
	\hline
		Node
	&	S
	&	pf
	&	Type
	\\
	\hline
		N05\phantom{-1}
	&	-20.6~kW
	&	0.97
	&   P/Q
	\\
		N09-1
	&	\phantom{-}49.1~kW
	&	0.95
	&   P/Q
	\\
		N09-2
	&	\phantom{-0}2.0~kW
	&	1.00
	&   P/Q
	\\
		N11-1
	&	\phantom{-}11.2~kW   
	&	0.95
	&   P/Q
	\\
		N11-2
	&	\phantom{-0}9.1~kW   
	&	0.95
	&   P/Q
	\\
		N13-1
	&	\phantom{-}10.5~kW   
	&	0.95
	&   P/Q
	\\
		N13-2
	&	-10.0~kW   
	&	1.00
	&   P/Q
	\\
		N03\phantom{-1}
	&	-20.0~kW   
	&	1.00
	&   Z
	\\
		N14\phantom{-1}
	&	-15.0~kW   
	&	1.00
	&   Z
	\\
		N23\phantom{-1}
	&	\phantom{-0}5.0~kW   
	&	-
	&   I
	\\
		N25\phantom{-1}
	&	\phantom{-}10.0~kW   
	&	-
	&   I
	\\
		N26\phantom{-1}
	&	\phantom{-0}5.0~kW   
	&	-
	&   I
	\\
		N24\phantom{-1}
	&	\phantom{0}-8.0~kW   
	&	-
	&   Z
	\\
	\hline
\end{tabular}
}
\end{table}

  

\begin{table}[t]
    \centering
    \caption{Short-Circuit Parameters of the Th{\'e}venin Equivalent.}
    \label{tab:TE:parameters}
    {

\renewcommand{\arraystretch}{1.1}

\begin{tabular}{ccl}
    \hline
        Parameter
    &   Value
    &   Description
    \\
    \hline
        $V_{n}$
    &   230\,V-\RMS
    &   Nominal voltage 
    \\
        $S_{\mathit{sc}}$
    &   630\,kW
    &   Short-circuit power
    \\
        $\Abs{Z_{\mathit{sc}}}$
    &   16.3\,m$\Omega$
    &   Short-circuit impedance
    \\
        $R_{\mathit{sc}}/X_{\mathit{sc}}$
    &   0.125
    &   Resistance-to-reactance ratio
    \\
    \hline
\end{tabular}
}
\end{table}

\begin{table}[t]
	\centering
	\caption{Harmonic Voltages of the Th{\'e}venin Equivalent (see \cite{Std:BSI-EN-50160:2000}).}
	\label{tab:TE:harmonics}
	{
\renewcommand{\arraystretch}{1.1}
\begin{tabular}{ccr}
    \hline
        $h$
    &   $|V_{\TE,h}|$
    &   \multicolumn{1}{c}{$\angle V_{\TE,h}$}
    \\
    \hline
        1
    &   1.0\,p.u.
    &   0\,rad
    \\
        5
    &   6.0\,\%
    &   $\pi$/8\,rad
    \\
        7
    &   5.0\,\%
    &   $\pi$/12\,rad
    \\ 
        11
    &   3.5\,\%
    &   $\pi$/16\,rad
    \\
        13
    &   3.0\,\%
    &   $\pi$/8\,rad
    \\
        17
    &   2.0\%
    &   $\pi$/12\,rad
    \\
        19
    &   1.5\,\%
    &   $\pi$/16\,rad
    \\
        23
    &   1.5\,\%
    &   $\pi$/16\,rad
    \\
    \hline
\end{tabular}
}

\end{table}

  

\begin{table}[htbp]
    \centering
    \caption{\AC ($+/-/0$) and \DC Parameters of the Lines.}
    \label{tab:grid:parameters}
	{
\renewcommand{\arraystretch}{1.1}
\setlength{\tabcolsep}{0.15cm}

\begin{tabular}{cccc}
    \hline
        ID
    &   $R_{+/-/0}$ or $R$ 
    &   $L_{+/-/0}$ or $L$ 
    &   $C_{+/-/0}$ or $C$ 
    \\
    \hline
        AC-T1
    &   3.30~$\Omega$/km
    &   0.45~mH/km
    &   150~nF/km
    \\
        AC-T2
    &   1.21~$\Omega$/km
    &   0.42~mH/km
    &   230~nF/km
    \\
        AC-T3
    &   0.78~$\Omega$/km
    &   0.40~mH/km
    &   210~nF/km
    \\
        AC-T4
    &   0.55~$\Omega$/km
    &   0.39~mH/km
    &   260~nF/km
    \\
        AC-T5
    &   0.27~$\Omega$/km
    &   0.38~mH/km
    &   320~nF/km
    \\
        DC-T1
    &   0.08~$\Omega$/km
    &   0.28~mH/km
    &   292~nF/km
    \\
    \hline
\end{tabular}
}

\end{table}



To assess the accuracy of the \HPF method compared to the \TDS so-called \emph{Key Performance Indicators} (\KPI[s]) are defined.
More specifically, the errors of the harmonic phasors between the \DFT of the \TDS and the results of the \HPF are derived.
Let the Fourier coefficient of a three-phase electrical quantity (i.e., voltage or current) be denoted as $\mathbf{X}_{h}$.
Then, the \KPI[s] are defined as:
\begin{align}
                e_{\textup{abs}}(\mathbf{X}_{h})
    &\coloneqq  \max_{p} \Abs{\Abs{X_{h,p,\HPF}}-\Abs{X_{h,p,\TDS}} }\\
                e_{\textup{arg}}(\mathbf{X}_{h})
    &\coloneqq  \max_{p} \Abs{ \angle X_{h,p,\HPF}- \angle X_{h,p,\TDS} }
\end{align}
In short, $e_{\textup{abs}}(\mathbf{X}_{h})$ and $e_{\textup{arg}}(\mathbf{X}_{h})$ represent the maximum absolute errors over all phases $p\in\phases$ in magnitude and angle, respectively.


\subsection{Results and Discussions}

\begin{figure}[htbp]
    \centering
    \subfloat[]
    {%
        \centering
        \includegraphics[width=1\linewidth]{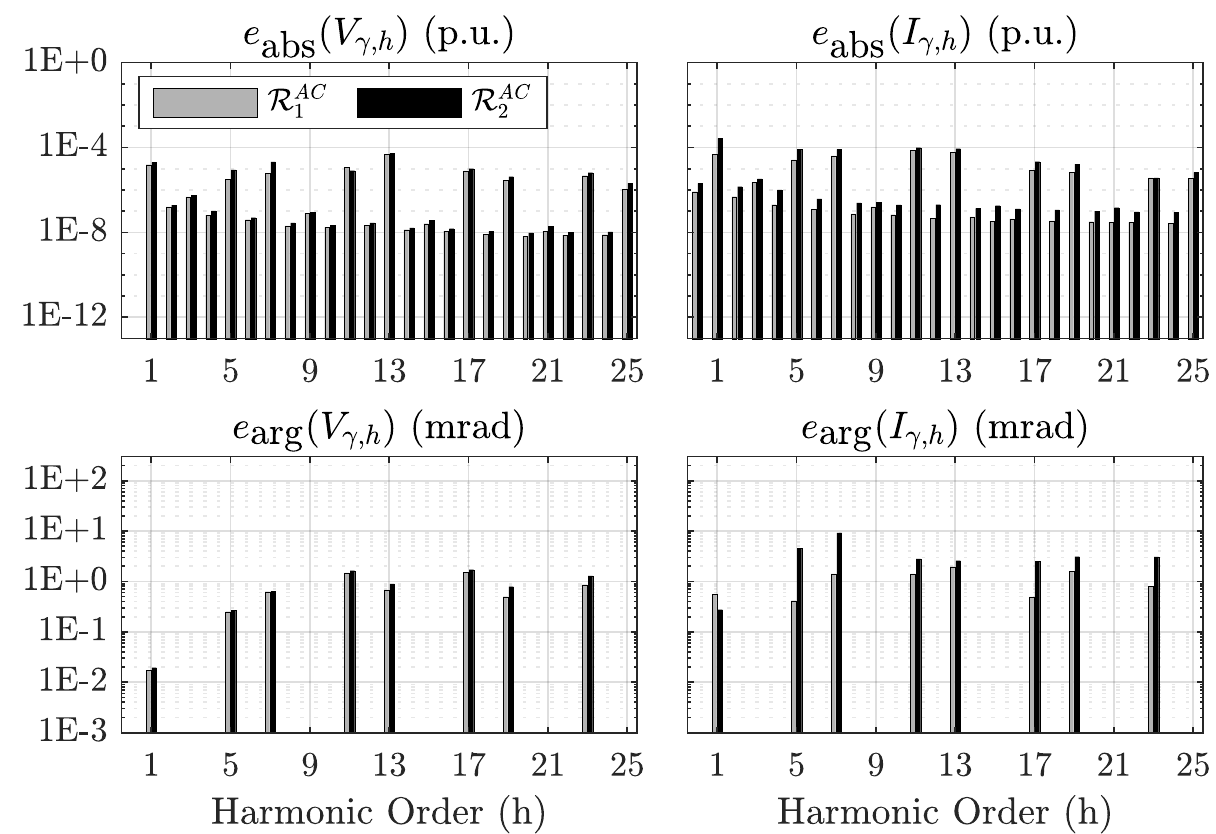}
        \label{fig:system:error:AC}
    }
    
    \subfloat[]
    {%
        \centering
        \includegraphics[width=1\linewidth]{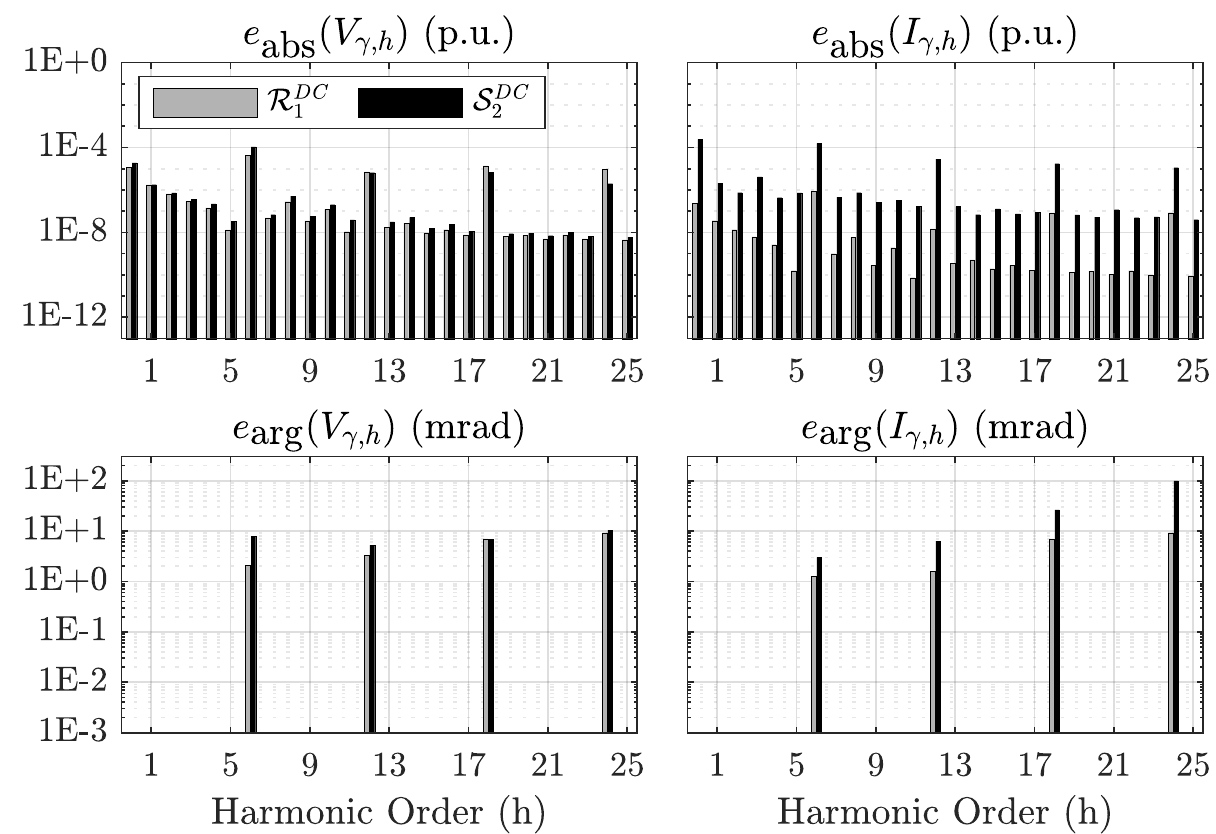}
        \label{fig:system:error:DC}
    }
    \caption
    {%
        Results of the validation of the \AC (\cref{fig:system:error:AC}) and \DC subsystem (\cref{fig:system:error:DC}).
        The grid-following \CIDER[s] ($\followers^{\AC}_1$) and the \AC side of the \NIC[s] ($\followers^{\AC}_2$), as well as the \DC-side resources ($\followers^{\DC}_1$) and the \DC side of the \NIC[s] ($\formers^{\DC}_2$) are shown.
        The plots show the maximum absolute errors over all nodes and phases, for voltages (left column) and currents (right column), in magnitude (top row) and phase (bottom row).
    }
    \label{fig:system:error}
\end{figure}

In \cref{fig:system:error} the accuracy of the \HPF algorithm for hybrid \AC/\DC grids is compared to the \TDS in Simulink.
The highest errors at every harmonic frequency and over all nodes and phases for nodal voltages and injected currents are shown for the two subsystems independently.
In \cref{fig:system:error:AC} the errors are shown for the set of all \CIDER[s] and \NIC[s], as well as the passive impedance loads.
For the \DC subsystem, the set of grid-forming (i.e., including the \DC quantities of the \NIC[s]) and grid-following nodes is shown.
The maximum errors regarding the voltages and currents occur in both subsystems in the nodes where the \NIC[s] are connected.
More precisely, the maximum errors in magnitude and phase for the voltages are $e_{\textup{abs}}(\mathbf{V}_{13})=5.37$E-5~p.u. and $e_{\textup{arg}}(\mathbf{V}_{17})=1.7$~mrad in the \AC subsystem and $e_{\textup{abs}}(\mathbf{V}_{6})=1.05$E-4~p.u. and $e_{\textup{arg}}(\mathbf{V}_{24})=10.5$~mrad in the \DC subsystem, respectively. 
The maximum errors w.r.t. current magnitude and angle are $e_{\textup{abs}}(\mathbf{I}_{1})=2.71$E-4~p.u. and $e_{\textup{arg}}(\mathbf{I}_{7})=9.2$~mrad in the \AC subsystem and $e_{\textup{abs}}(\mathbf{I}_{0})=2.38$E-4~p.u. and $e_{\textup{arg}}(\mathbf{I}_{24})=99.5$~mrad in the \DC subsystem, respectively.
The obtained errors for both subsystems are lower than the accuracy of standard measurement equipment (i.e., they are unobservable in practice).
Thus, the accuracy of the \HPF method for hybrid \AC/\DC grids is validated.

\section{Conclusions}
\label{sec:Conclusions}

This paper proposes an \HPF algorithm for the analysis of hybrid \AC/\DC grids, which is an extension of the \HPF algorithm for \AC grids introduced in \cite{jrn:2020:kettner-becker:HPF-1,jrn:2020:kettner-becker:HPF-2,jrn:2022:becker}.
In hybrid grids, the \AC and \DC subsystems are interconnected through \NIC[s].
Those \NIC[s] have two ports: i.e., one on the \AC side and one on the \DC side.
To include this behaviour into the previously developed \HPF framework, resources are separated into single-port and two-port devices.
Single-port resources (e.g., \CIDER[s]) lie within one single subsystem, while two-port resources (i.e., \NIC[s]) interconnect two different subsystems.
The \HPF algorithm is formulated based on the mismatch equations of the nodal equations as seen from the grid and the resources, respectively, and solved via the Newton-Raphson method.
In order to include the \NIC[s] and the \DC subsystem, the structure of the Jacobian matrix of the \HPF is suitably modified.
More precisely, it is written as the difference between the Jacobian matrices associated with the grid and the resource equations, respectively.
The former matrix has a block-diagonal structure, and exhibits no coupling between \AC and \DC quantities.
The latter matrix does feature coupling terms (i.e., non-zero off-diagonal blocks), which are due to the two-port nature of the \NIC[s].
The \HPF method for hybrid \AC/\DC grids is validated through \TDS in Simulink.
The errors observed in these simulations demonstrate the high accuracy of the \HPF w.r.t. the \TDS.
The largest observed errors are $2.71$E-4~p.u. w.r.t. current magnitude, $1.05$E-4~p.u. w.r.t. voltage magnitude, and 99.5~mrad w.r.t. phase (at the edge of the modelled frequency window).
This demonstrates that the proposed \HPF method for hybrid \AC/\DC grids can accurately analyze the interaction of harmonics between entire \AC and \DC subsystems.



\bibliographystyle{IEEEtran}
\bibliography{Bibliography}

\end{document}